\documentclass[conference]{IEEEtran}
\usepackage{amsmath,amssymb}
\usepackage{comment}
\usepackage{amssymb}
\usepackage{bm}
\usepackage[dvipdfmx]{graphicx}

\usepackage[ruled]{algorithm}
\usepackage{algorithmic}

\graphicspath{{figs/}}

\usepackage{setspace}

\usepackage{algorithm}
\usepackage{algorithmic}

\begin{document}
\title{Evaluation of Symmetric Mutual Information of the Simplified TDMR Channel Model} 

\author{
  \IEEEauthorblockN{Tadashi Wadayama}
  \IEEEauthorblockA{Department Computer Science and Engineering, \\Nagoya Institute of Technology,
    Nagoya,  Japan\\
    Email: wadayama@nitech.ac.jp} 
}

\maketitle
\begin{abstract}
In the present paper, a simplified two-dimensional magnetic recording (TDMR) channel model is proposed in order to capture the qualitative features of writing and read-back processes of TDMR systems.
The proposed channel model incorporates the effects of both linear interference from adjacent bit-cells and signal-dependent noise due to irregular grain boundaries between adjacent bit-cells. 
The simplicity of the proposed model enables us to derive the closed form of the conditional PDF representing the probabilistic nature of the channel. The conditional PDF is Gaussian distributed and is parameterized by a signal-dependent covariance matrix. Based on this conditional PDF, a Monte Carlo method for approximating the symmetric mutual information of this channel is developed. The symmetric mutual information is closely related to the areal density limit for TDMR systems. The numerical results suggest that we may need low-rate coding, e.g., 2/3 or 1/2, when the jitter-like noise becomes dominant.
\end{abstract}

\section{Introduction}

Continuing demand for high-density magnetic recording 
requires novel technology for realizing higher areal density.
The emergence of shingled writing promotes research 
into advanced signal processing because narrowing inter-track pitch causes 
severe inter-track interference and jitter-like noise around track edges. 
Under such circumstances, two-dimensional signal processing based on 
multiple reading heads is becoming a hot research topic in this field.
For example, such a system requires linear interference cancellation in both the cross-track and down-track directions.

In the near future, the size of bit-cells will likely shrink to be comparable 
to grains of a magnetizing material. As such, 
highly advanced two-dimensional signal processing will be required 
to handle strong linear interference from adjacent bit-cells and
signal-dependent noise due to irregular grain boundary between adjacent cells.
The concept of two-dimensional magnetic recording (TDMR) advocated by Wood et al. \cite{Wood}
has inspired research in this field. Related research into, for example,
signal processing \cite{Hwang}, detection algorithms \cite{Khatami}, and
simplified channel models \cite{Krishnan} has been conducted for TDMR systems.

One goal of the present paper is to develop a simple mathematical model 
for TDMR systems that is useful for
performance evaluation of detection algorithms 
and for the design of two-dimensional codes.
The {\em simplified TDMR channel model} presented herein 
incorporates the effects of both linear interference from adjacent bit-cells and  
signal-dependent noise. Although the channel model is fairly simple,
the model shows the complicated nature of signal-dependent channels.
Observing the behavior of the channel model 
will help to clarify the characteristics of TDMR channels.
The second goal of the present paper is to present a method for evaluating 
the symmetric mutual information. The evaluation of mutual information is 
not a trivial problem for a channel with signal-dependent noise, but 
the problem is worth pursuing because it is closely related to 
the areal density limit of TDMR systems.

\section{Two bit-cell model}

Before introducing the simplified TDMR channel model, we will discuss a 
two-bit-cell model in this section. Although this channel model is simple, it provides insight for a TDMR channel and this model has a number of features 
that a TDMR channel model must possess. For example,
this channel model can handle signal-dependent noise generated by the irregular boundary between 
two adjacent bit-cells.

A number of grains of a magnetizing material 
are spread over the surface of the recording medium. 
In the writing phase, a writing head magnetizes each grain upward or downward, depending on 
the information to be written.
A magnetization process of each grain is determined 
according to the strength distribution of magnetic flux emitted from a writing head and
the magnetic sensitivity of the grain. The first approximation introduced in the present paper is that 
the head footprint of magnetic flux has an exactly rectangular shape.
Namely, the magnetic field induced by a writing head has an effect on 
the rectangular area.
This approximation helps to simplify the channel modeling process described below.
This rectangular area is referred to herein as a {\em bit-cell}.
Some of the grains lie between two adjacent bit-cells and cause 
irregular grain boundary between adjacent bit-cells. 
Since a read-back system has no knowledge on the positions of grains, 
such an irregular grain boundary incurs jitter-like signal-dependent noise that makes it 
difficult to achieve reliable detection in TDMR systems.

A read-back signal is obtained by taking the convolution of 
a sensitivity function (i.e., 2D-impulse response) of a reading head 
and a magnetic field pattern caused by magnetized grains on the surface of 
a recording medium. A sample sequence of the read-back signals is sent to 
a signal processing and detection unit to estimate the written data.
In a TDMR system, two adjacent bit-cells in a grid of bit-cells are closely arranged 
and a reading head with a wide (compared to the size of a bit-cell) footprint is 
often considered in a TDMR system.
Two-dimensional linear interference consisting of both inter-symbol interference (ISI) and inter-track interference (ITI)
arises in such a system and should be appropriately processed to provide 
high reliability of estimated data.

\subsection{Details}

\begin{figure}
\begin{center}
\includegraphics[scale=0.5]{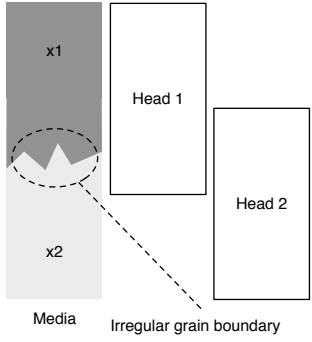}
\end{center}
\caption{Two bit-cell model}
\label{tbmodel}
\end{figure}

The two-bit-cell model, which is an abstraction of a part of a TDMR system, 
is shown in Fig.~\ref{tbmodel}. This model contains two adjacent bit-cells, and
the information written to these bit-cells is denoted by $x_1, x_2 \in \{+1, -1\}$.
Between two bit-cells, we assume that an irregular grain boundary exists.
We also assume that two overlapping footprints (denoted by Head 1 and Head 2 in Fig.~\ref{tbmodel}) 
of a reading head are exploited to yield the two read-back signals.

In the two-bit-cell model, the two read-back signals $y_1$  and $y_2$
are given by
\begin{eqnarray} \nonumber
y_1 &=& x_1 + \left(\frac 1 2 + n_j \right) x_1 + \left(\frac 1 2  - n_j \right) x_2 + n_1 \\ \label{twobit}
y_2 &=& x_2 + \left(\frac 1 2- n_j \right) x_2 + \left(\frac 1 2 + n_j \right) x_1 + n_2.
\end{eqnarray}
Note that the read-back signals are assumed to be appropriately scaled so as not to introduce 
additional scale parameters.
The symbols $n_1$ and $n_2$ are  independent Gaussian random variables
with mean $0$ and variance $\sigma_s^2$ that represent system noises.
The symbol $n_j$ is an independent Gaussian random variable 
with mean $0$ and variance $\sigma_j^2$ that corresponds
to the irregular grain boundary between two bit-cells. In other words,
we introduce a stochastic model for expressing the effects of an irregular grain boundary.

The following is an implicit assumption to derive the above model of the read-back signals.
Let $f(a,b)$ be the vertical component of magnetic flux induced by the media at 
point $(a,b) \in \Bbb R$. For simplicity, it is assumed that $f(a,b) = x_1 \in \{+1,-1\}$ if 
$(a,b)$ is in the magnetized area corresponding to bit-cell 1; otherwise, 
$f(a,b) = x_2\in \{+1,-1\}$ is assumed. Under this setting, the read-back signal 
$y_i$ is obtained by
\begin{equation}
y_i = \int_{Head\  i} f(a,b) \ da\  db.
\end{equation}
A read-back signal can be considered as an output from 
a concatenated system of a 2-D low-pass filter with a rectangular window (i.e., head footprint)
followed by a 2-D sampler.  Although this assumption 
is based on an ideal case (such a 2-D low-pass 
filter is not realizable), we expect the model to reflect the qualitative nature of an actual TDMR system. 

If there is no system noise or jitter-like noise due to the irregular grain boundary between two bit-cells, the following read-back signals can be obtained:
\begin{eqnarray} \nonumber
y_1 &=& \frac 3 2 x_1 + \frac 1 2  x_2,\quad 
y_2 = \frac 3 2 x_2 + \frac 1 2  x_1.
\end{eqnarray}
Some of the grains located in the middle of the bit-cell boundary have the same polarity as the value of $x_1$, whereas 
others have the polarity same as the value of $x_2$. The balance of contributions of 
such grains are modeled by a Gaussian random variable $n_j$ in the proposed model.
We adopt a Gaussian model for representing this balance for two reasons:
1) as described later herein, a Gaussian model leads to a mathematically tractable probability density function (PDF) of
received signals, and 2) since the balance of areas $n_j$ consists of  contributions from several grains
(i.e., addition of several random variables), it is natural to model the balance of areas $n_j$ by a Gaussian random variable.

We can rewrite the model (\ref{twobit}) in vector form, as follows:
\begin{eqnarray} \nonumber
\left(
\begin{array}{c}
y_1 \\
y_2 \\
\end{array}
\right)
&=& 
\frac 1 2 
\left(
\begin{array}{cc}
3 & 1\\
1  & 3 \\
\end{array}
\right)
\left(
\begin{array}{c}
x_1 \\
x_2 \\
\end{array}
\right)
+
\left(
\begin{array}{c}
n_1 \\
n_2 \\
\end{array}
\right) \\ 
&+& 
n_j 
\left(
\begin{array}{cc}
1 & -1\\
1  & -1 \\
\end{array}
\right)
\left(
\begin{array}{c}
x_1 \\
x_2 \\
\end{array}
\right).
\end{eqnarray}
The first term on the right-hand side of the above equation represents a linear interference that occurs on $x_1$ and $x_2$, 
and the second term on the right-hand side is the system noise.
The third term on the right-hand side indicates the jitter-like noise due to the irregular grain boundary.
Note that the last term on the right-hand side disappears if $x_1 = x_2$.
Namely, when two bit-cells have the same polarity, an irregular grain boundary 
has no harmful effect regarding detection. The last term on the right-hand side, representing
the signal-dependent noise, introduces a certain difficulty in 
designing detection algorithms, codes, and evaluating the capacity of the channel.

\begin{figure}
\begin{center}
\includegraphics[scale=0.4]{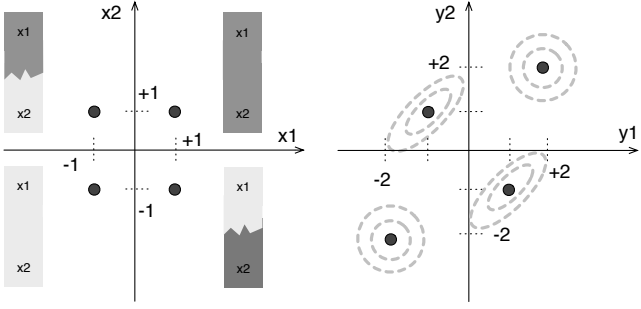}
\end{center}
\caption{Signal constellations (left: transmitted symbols $(x_1,x_2)$, 
right: received symbols $(y_1,y_2)$ in a noiseless case)}
\label{constelation}
\end{figure}

Figure \ref{constelation} indicates signal constellations regarding the two-bit-cell model.
The transmitted symbols $(x_1,x_2)$ are depicted in Fig.~\ref{constelation} (left),
and the received symbols $(y_1,y_2)$ in a noiseless case are indicated as black dots in Fig.~\ref{constelation} (right).
Due to the linear interference, we can see that 
the received symbols are located at the corner points of a diamond-shaped region.
The dotted circles around the black dot represent the contours of the PDF $P_Y(y_1,y_2) (Y = (y_1, y_2))$.
The exact form of the PDF will be discussed in the next subsection.
Note that the shape of the contours around the dots depends on the signals.
For example, the noiseless received signals $(y_1,y_2) =(+2,+2)$ correspond to the
transmitted signals $(x_1,x_2) = (+1,+1)$. In this case, no jitter-like noise appears in the detection phase,
and only system noise $(n_1,n_2)$ disturbs the received signals. On the other hand, 
when $(x_1, x_2) = (+1,-1)$, the irregular grain boundary is as depicted in Fig.~\ref{constelation} (right).

\subsection{Conditional probability density function of received signals}

One of the advantages of the two-bit-cell model is its mathematical tractability. It can be shown that the conditional PDF of the received symbols are Gaussian distributed (details are discussed in the next section). The conditional PDF $P_{ \mathbf{Y}| \mathbf{X}}(\mathbf{y}| \mathbf{x})$ $(\mathbf{x}=(x_1,x_2)^T,  \mathbf{y}=(y_1,y_2)^T)$ is given by
\begin{eqnarray} \nonumber
P_{ \mathbf{Y}| \mathbf{X}}(\mathbf{y}| \mathbf{x}) 
\hspace{-3mm}&=& \hspace{-3mm}
\frac{1}
{2 \pi  \sqrt{|S(\mathbf{x})| } }
 \exp
   \left(-
   (\mathbf{y}-A \mathbf{x})^T 
S(\mathbf{x})^{-1}
(\mathbf{y}-A \mathbf{x})
   \right),
\end{eqnarray}
where the matrix $A$ is the interference coefficient matrix defined as 
\begin{equation}
A = \frac 1 2 \left(
\begin{array}{cc}
3 & 1 \\
1 & 3
\end{array}
\right).
\end{equation}
The covariance matrix $S(\mathbf{x})$ is given by
\begin{equation}
S(\mathbf{x}) = \left(
\begin{array}{cc}
 \sigma_j^2 (x_1-x_2)^2 + \sigma_s^2 &  \sigma_j^2 (x_1-x_2)^2 \\
 \sigma_j^2 (x_1-x_2)^2 & \sigma_j^2 (x_1-x_2)^2 + \sigma_s^2 
\end{array}
\right).
\end{equation}

From $P_{ \mathbf{Y}| \mathbf{X}}(\mathbf{y}| \mathbf{x})$, 
the noise vector $\mathbf{y}-A \mathbf{x}$ is Gaussian distributed, 
and the covariance matrix $S(\mathbf{x})$ is dependent on the transmitted signal $\mathbf{x}$. This observation matches the contours shown in Fig.~\ref{constelation} (right).

\subsection{Maximum likelihood detection}

We have had the exact conditional PDF for the two-bit-cell model.
The conditional PDF  leads to  the explicit formula for maximum likelihood detection for 
this channel. At a glance, the two-bit-cell model appears to be a simple model, but 
this model possesses properties that are not so intuitive. In this subsection, we will observe such 
properties in terms of signal detection.

Let us define a distance measure $D(\mathbf{y}| \mathbf{x})$  as
\begin{equation}
D(\mathbf{y}| \mathbf{x}) =  (\mathbf{y}-A \mathbf{x})^T 
S(\mathbf{x})^{-1}
(\mathbf{y}-A \mathbf{x}) + \log|S(\mathbf{x})|,
\end{equation}
which is derived by taking the negative logarithm to the conditional PDF of the channel.
Using this distance measure, the maximum likelihood (ML) detection rule can be written as
\begin{equation}
\hat{\mathbf{x} } = \arg_{\mathbf{x} \in \{+1,-1\}^2 } \min D(\mathbf{y}| \mathbf{x}).
\end{equation}

\begin{figure}[htbp]
\begin{center}
\includegraphics[scale=0.7]{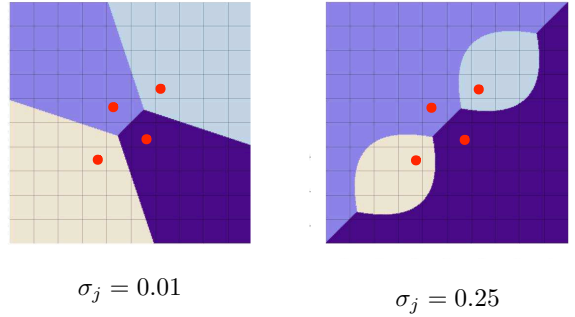}
\end{center}
\caption{ML decision region (horizontal and vertical axes represent $y_1 (-6 \le y_1 \le +6)$ and $y_2 (-6 \le y_2 \le +6)$, respectively. 
The red dot indicates the received points in the noiseless case.)}
\label{mlregion}
\end{figure}

Figure \ref{mlregion} shows the decision regions of the two-bit-cell model for this ML detection rule in the $(y_1, y_2)$ plane. From left to right, the cases of standard deviations $\sigma_j = 0.01$ and $0.25$ are depicted. Unlike a conventional linear interference channel, we can see that two decision regions for $(y_1,y_2)=(+1,+1),(-1,-1)$ become closed sets as the standard deviation $\sigma_j$ increases. The peculiar shape of these decision regions suggests the difficulty of a direction problem for a channel model containing a number of bit-cells.
Another observation we can make from this figure is that we may need a non-linear 
detector instead of a linear detector in order to approximate the exact ML detector with 
reasonable computational complexity.

\section{Simplified TDMR channel model}

In this section, we generalize the two-bit-cell model discussed in the previous section 
into a more general two-dimensional channel model that is referred to as 
a simplified TDMR channel model.

\begin{figure}[htbp]
\begin{center}
\includegraphics[scale=0.3]{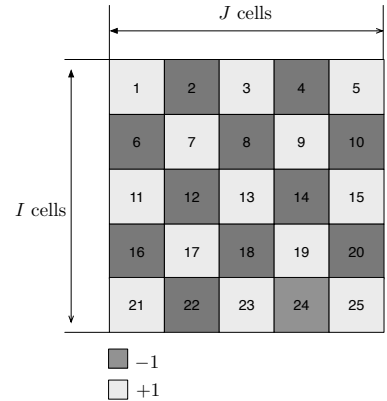}
\end{center}
\caption{Writable area and bit-cells ($n = 25$)}
\label{bitcell}
\end{figure}

In the following discussion, we will assume that bit-cells are arranged on grid points in a rectangular area, as shown in Fig. \ref{bitcell}. The area of interest is called the {\em writable region} that consists of $I$ rows and $J$ columns. Each bit-cell has its index $i (i \in [1,n], n = IJ)$, and the bit-cell with index $i$ is denoted by $B_i$.
As in the case of the two-bit-cell model, each bit-cell can take the value $+1$ or $-1$.
The notation $[a,b]$ represents the set of consecutive integers from $a$ to $b$.

\begin{figure}[htbp]
\begin{center}
\includegraphics[scale=0.55]{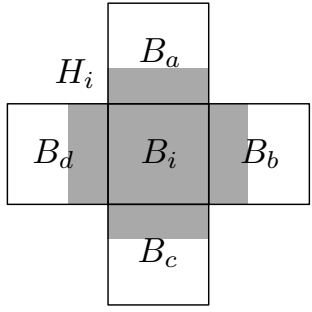}
\end{center}
\caption{Head footprint $H_{i}$ (gray area corresponds to the head footprint $H_i$)}
\label{head}
\end{figure}

The set of indices of bit-cells adjacent to the bit-cell $B_i$ is denoted by
$N(i) (i \in [1,n])$. For example, in the case of Fig. \ref{bitcell}, we have $N(13)=\{8,12,14,18 \}$. The boundary between two adjacent bit-cells is referred to as an {\em edge}. The edge between bit-cells $B_i$ and $B_j$ is represented by $E_{i,j}$, or equivalently $E_{j,i}$.
The {\em head footprint} $H_{i}$ corresponding to bit-cell $B_i(i \in [1,n])$
provides a received symbol $y_i$ by reading a certain area around $B_i$.
Here, we assume that the head footprint has a crisscross shape, as shown in Fig.~\ref{head}.

In the simplified TDMR channel model,
the received signal $y_i (i \in [1,n])$ corresponding to the head footprint $H_{i}$
is modeled by
\begin{equation} \label{model}
y_{i} 
= \alpha x_i + \beta \sum_{j \in N(i)} x_j + \sum_{j \in N(i) }|x_i - x_j | z_{i,j}  + z_i.
\end{equation}
The real constants $\alpha$ and $\beta$ are scale parameters for the intended signal and 
the linear interference coming from adjacent cells, respectively.
The noise term $z_i$ is a Gaussian 
i.i.d.\ random variable with mean 0 and variance $\sigma_s^2$ and 
represents a system noise.
An i.i.d.\ random variable $z_{i,j}$ is identical to $z_{j,i}$
and represents jitter-like noise due to the irregular grain boundary around edge $E_{i,j}$.
The random variable $z_{i,j}$ follows the one-dimensional Gaussian PDF with mean $0$ and variance $\sigma_j^2$.
Of course, it is easy to provide more flexibility by introducing more parameters. For example,
the coefficients of linear interference from the adjacent cells can be direction dependent. However,
we will avoid introducing too many parameters so as not to complicate the following discussion.
Once the following argument is grasped, modification of the model would be straightforward.

\begin{figure}[htbp]
\begin{center}
\includegraphics[scale=0.6]{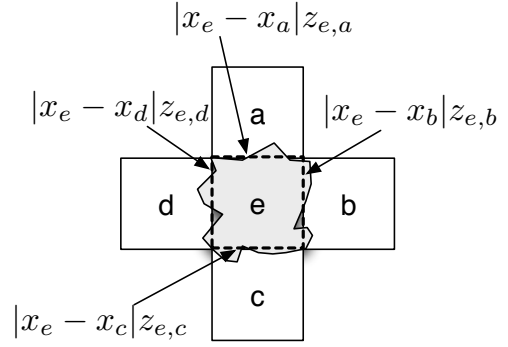}
\end{center}
\caption{Expressing irregular grain boundary with Gaussian random variable}
\label{randedge}
\end{figure}

As in the case of the two-bit-cell model, jitter-like noise occurs around 
an edge between two bit-cells having distinct polarities (Fig.~\ref{randedge}).
When the size of a bit-cell becomes comparable to the average diameter of grains,
the effect of an irregular grain boundary around an edge tends to be the dominant source of noise 
that degrades the overall detection performance of a system. The proposed model includes the effect of the irregular grain boundary by introducing random variables corresponding to each edge in the writable area. If $x_i$ and $x_j$ have the same value (i.e., same polarity), the term expressing the jitter-like noise $|x_i - x_j| z_{i,j}$ disappears. On the other hand, when $x_i$ and $x_j$ have opposite polarities, the term $|x_i - x_j| z_{i,j}$ influences the received signal $y_i$ as a signal-dependent noise.

For the following analysis, it is convenient to introduce a vector notation 
$
\mathbf{y} = A \mathbf{x} + \mathbf{q}(\mathbf{x})
$
to express the model (\ref{model}) of the received signals, where 
\begin{equation}
\mathbf{x} = (x_1,x_2,\ldots, x_n)^T,\  \mathbf{y} = (y_1,y_2,\ldots, y_n)^T.
\end{equation}
The matrix $A \in \Bbb R^{n \times n}$ represents the linear interference coefficients
and $\mathbf{q}(\mathbf{x})$ denotes the noise term
\begin{equation}
\mathbf{q}(\mathbf{x}) = (q_1(\mathbf{x}), q_2(\mathbf{x}),\ldots, q_n(\mathbf{x})),
\end{equation}
where the explicit form of $q_i(\mathbf{x})$ is given by 
\begin{equation}
q_i(\mathbf{x}) =  \sum_{j \in N(i) }|x_i - x_j | z_{i,j}  + z_i.
\end{equation}

The first and second moments of the noise term $q_i(\mathbf{x})(i \in [1,n])$ can be immediately derived 
as follows.  The second moment of $q_i(\mathbf{x})$ is evaluated as
\begin{eqnarray}
E[q_i^2(\mathbf{x})] &=& \sum_{k \in N(i) } (x_i - x_k)^2 E[z_{i,k}^2] + E[z_i^2] \\
&=& \sum_{k \in N(i) } (x_i - x_k)^2 \sigma_j^2 + \sigma_s^2. 
\end{eqnarray}
In the above derivation, the independences of $z_{i,j}$ and $z_i$  are exploited.
In a similar manner, the second cross moment $E[q_i(\mathbf{x})q_j(\mathbf{x})] (i \ne j)$ is given by
\begin{eqnarray} \nonumber
E[q_i(\mathbf{x})q_j(\mathbf{x})] 
&=&  
\left\{
\begin{array}{ll}
(x_i - x_j)^2  \sigma_j^2, & \mbox{$B_i$ and $B_j$ are adjacent} \\
0, &  \mbox{otherwise}.
\end{array}
\right.
\end{eqnarray}
In summary, the covariance matrix regarding $\mathbf{q}(\mathbf{x})$, 
which is denoted by $S(\mathbf{x}) = \{S_{i,j}(\mathbf{x}) \}$,  is given by
\begin{equation}
S_{i,j}(\mathbf{x}) = 
\left\{
\begin{array}{ll}
\sum_{k \in N(i) } (x_i - x_k)^2 \sigma_j^2 + \sigma_s^2,\ &  i = j \\
(x_i - x_j)^2  \sigma_j^2,\ &  j \in N(i)  \\
0, &  \mbox{otherwise}.
\end{array}
\right.
\end{equation}

The vector $\mathbf{q}(\mathbf{x})$ can be obtained by multiplying a real matrix to a vector of independent Gaussian random variables. 
This implies that $\mathbf{q}(\mathbf{x})$ obeys the multi-dimensional Gaussian PDF.
We therefore have the conditional PDF given by
\begin{eqnarray}  \nonumber
P_{\mathbf{Y}|\mathbf{X}}(\mathbf{y}| \mathbf{x})
\hspace{-3mm}
&=&
\hspace{-3mm}
 \frac{1}{(\sqrt{2 \pi}  )^m \sqrt{|S(\mathbf{x})|} }  \\ \label{condpdf}
&\times& \hspace{-3mm}\exp \left(- \frac 1 2 (\mathbf{y} - A \mathbf{x})^T S^{-1}(\mathbf{x})(\mathbf{y} - A \mathbf{x})  \right).
\end{eqnarray}
It is trivial to see that the conditional PDF of the two-bit-cell model is a special case of this 
conditional PDF (\ref{condpdf}).

As an example, let us consider the case in which the writing area consists of $2 \times 2$ bit-cells.
When all four bit-cells have the same polarity,  the corresponding covariance matrix 
becomes a diagonal matrix with the diagonal element $\sigma_s^2$
because no jitter-like noise occurs.
The worst pattern in terms of noise variance is the checker-board-like pattern 
(i.e., any pair of adjacent bit-cells have distinct polarity) because 
every edge between two bits causes an irregular grain boundary effect.
In such a case, the covariance matrix has the following form:
\begin{equation}
 \left(
\begin{array}{cccc}
\sigma_s^2+8 \sigma_j^2 & 4\sigma_j^2 & 0 & 4\sigma_j^2 \\
4\sigma_j^2 & \sigma_s^2+8 \sigma_j^2 &4\sigma_j^2 & 0 \\
0& 4\sigma_j^2 & \sigma_s^2+8 \sigma_j^2 & 4\sigma_j^2  \\
4\sigma_j^2 & 0& 4\sigma_j^2 & \sigma_s^2+8 \sigma_j^2   \\
\end{array}
\right).
\end{equation}

\section{Evaluation of symmetric mutual information}

Evaluation of the capacity of the simplified TDMR channel is of practical importance. 
The capacity indicates the {\em areal density limit} for the TDMR channel without making any assumptions on coding and detection methods 
and may also give us insight into the design of a read-back system including detection algorithms and coding.

In the previous section, we have derived the closed form expression of the conditional PDF
that perfectly characterizes the channel model. Therefore, in principle, we have sufficient information to evaluate the capacity of the channel given as
$
C = \max_{P_{\mathbf{X}}} I(\mathbf{X}; \mathbf{Y}).
$
The random variables  $\mathbf{X} = (X_1, X_2,\ldots, X_n)^T, \mathbf{Y} = (Y_1, Y_2,\ldots, Y_n)^T$
represent written signals in a writable area and read-back signals, respectively.
The symbol $I(\mathbf{X}; \mathbf{Y})$ is the mutual information defined by
\begin{equation}
I(\mathbf{X}; \mathbf{Y}) = H(\mathbf{Y}) - H(\mathbf{Y} | \mathbf{X}).
\end{equation}
However, the problem of evaluating the capacity of this channel is not trivial.
Of course, maximization of mutual information $I(\mathbf{X}; \mathbf{Y})$
in terms of the prior distribution $P_{\mathbf{X}}$
is a computationally difficult problem. Even the evaluation of the mutual information itself 
is not a simple problem because we need to handle signal-dependent noise that results in 
a nonuniform conditional PDF. 
Another computational difficulty comes from the intrinsically high dimensionality of the model. 
High-dimensional integrations are required for evaluating the mutual information.

In the present paper, we will focus on the symmetric mutual information instead of the channel capacity. 
The symmetric mutual information is the mutual information under the assumption that 
the written signal $\mathbf{X}$ is equiprobable. From the definition of the capacity, it is clear that 
the symmetric mutual information is a lower bound of the capacity and can be used as an 
approximate value of the capacity. This simplification makes the problem computationally tractable. 
In this section, we present a Monte Carlo method for evaluating the symmetric mutual information of the simplified TDMR channel model.

\subsection{Symmetric mutual information}

Under the assumption that $\mathbf{X}$ is equiprobable, 
the probability $P_{\mathbf{Y}}(\mathbf{y})$ is given by
\begin{eqnarray} \nonumber
P_{\mathbf{Y}}(\mathbf{y}) 
&=& \sum_{\mathbf{x} \in \{+1,-1 \}^n} P_{\mathbf{Y}|\mathbf{X}}(\mathbf{y}| \mathbf{x}) P_{\mathbf{X}}(\mathbf{x}) \\ \label{PY}
&=& 2^{-n} \sum_{\mathbf{x} \in \{+1,-1 \}^n} P_{\mathbf{Y}|\mathbf{X}}(\mathbf{y}| \mathbf{x}),
\end{eqnarray}
where  the conditional PDF $P_{\mathbf{Y}|\mathbf{X}}(\mathbf{y}| \mathbf{x})$ is given in (\ref{condpdf}).
The entropy of $\mathbf{Y}$ is thus given by the following multi-dimensional integration:
\begin{equation} 
H(\mathbf{Y}) = - \int_{\mathbf{y} \in \Bbb R^n} P_{\mathbf{Y}} (\mathbf{y}) \log_2 P_{\mathbf{Y}} (\mathbf{y}) d \mathbf{y}.
\end{equation}
The entropy $H$ of a multi-dimensional Gaussian random variable with 
a covariance matrix $K$  \cite{Cover} is given by
\begin{equation}
H = (1/2) \log_2 \left((2 \pi e)^n |K| \right).
\end{equation}
Using this formula, we can derive the conditional entropy $H(\mathbf{Y}| \mathbf{X})$
as follows:
\begin{eqnarray} \nonumber
H(\mathbf{Y}| \mathbf{X}) 
&=& - \sum_{\mathbf{x}} 
P_{\mathbf{X}}(\mathbf{x})  \int_{\mathbf{y}}  P_{\mathbf{Y}|\mathbf{X}}(\mathbf{y}|\mathbf{x})
 \log_2 P_{\mathbf{Y}| \mathbf{X}}( \mathbf{y}| \mathbf{x}) d \mathbf{y} \\ \label{condent}
&=& 2^{-n} \sum_{\mathbf{x}}  \frac 1 2 \log_2 \left((2 \pi e)^n |S(\mathbf{x}) | \right).
\end{eqnarray}
The range of summation $\mathbf{x} \in \{+1,-1\}^n $is hereafter omitted.
The symmetric mutual information, denoted by $I_S$,  is thus given by
\begin{eqnarray} \nonumber
&&\hspace{-8mm} I_S \\ \nonumber
&=&- \int_{\mathbf{y} \in \Bbb R^n} 
\left(\frac{\sum_{\mathbf{x} } P_{\mathbf{Y}|\mathbf{X}}(\mathbf{y}| \mathbf{x})}{2^n} \right) 
\log_2 \left(\frac{\sum_{\mathbf{x} } P_{\mathbf{Y}|\mathbf{X}}(\mathbf{y}| \mathbf{x})}{2^n} \right) d \mathbf{y} \\
&-& 2^{-n} \sum_{\mathbf{x} }  \frac 1 2 \log_2 \left((2 \pi e)^n |S(\mathbf{x}) | \right).
\end{eqnarray}

\subsection{Monte Carlo method for evaluation of $I_S$}

In order to circumvent the numerical difficulty of high-dimensional numerical integration, we use 
a Monte Carlo method to evaluate the symmetric mutual information.
The pseudo code of the Monte Carlo method is summarized as Algorithm 1.

\begin{algorithm}                      
\caption{Monte Carlo method for evaluation of $I_S$}          
\label{alg1}                          
\begin{algorithmic}[1]                
\STATE According to (\ref{condent}),  evaluate $H(\mathbf{Y}| \mathbf{X})$ exactly. 
\STATE $t:=0; s:=0$
\WHILE {$t < t_{max}$}
\STATE Generate $\mathbf{x}$ according to the uniform distribution. 
\STATE Generate a received vector $\mathbf{y} = A \mathbf{x} + q(\mathbf{x})$. 
\STATE According to (\ref{PY}),  evaluate $P_{\mathbf{Y}}(\mathbf{y})$ exactly.
\STATE $s := s  - \log_2 P_{\mathbf{Y}}(\mathbf{y})$
\STATE $t := t + 1$
\ENDWHILE
\STATE $\theta :=  s/t_{max}$
\STATE Output $\theta - H(\mathbf{Y}| \mathbf{X})$ as an estimate of $I_S$.
\end{algorithmic}
\end{algorithm}

The parameter $t_{max}$ is a given positive number that indicates the number of iterations.
In the following, we explain the important steps in Algorithm 1.
In line 1 in Algorithm 1,  $H(\mathbf{Y}| \mathbf{X})$ is evaluated according to (\ref{condent}).
Since the number of summands of (\ref{condent}) is $2^n$, this step requires time complexity $O(2^n)$. The step at line 6 is also a time consuming part of the algorithm that requires $O(2^n)$-time for evaluation. Note that the evaluation of $P_{\mathbf{Y}}(\mathbf{y})$ is repeated $t_{max}$ times.
Thus, this part dominates the overall computation time.
The quantity $\theta$ is an approximate value of $H(\mathbf{Y})$ as 
\[
\theta \simeq {\sf E}_{P_{\mathbf{Y}}} [ - \log_2 P_{\mathbf{Y}}(\mathbf{y})] = H(\mathbf{Y}).
\]
There is a tradeoff relationship between the computation complexity and accuracy 
of the approximation regarding $\theta$. The accuracy improves as the number of iterations 
$t_{max}$ increases but a larger $t_{max}$ results in a longer computation time.

\section{Numerical results}

\begin{figure}[htbp]
\begin{center}
\includegraphics[scale=0.7]{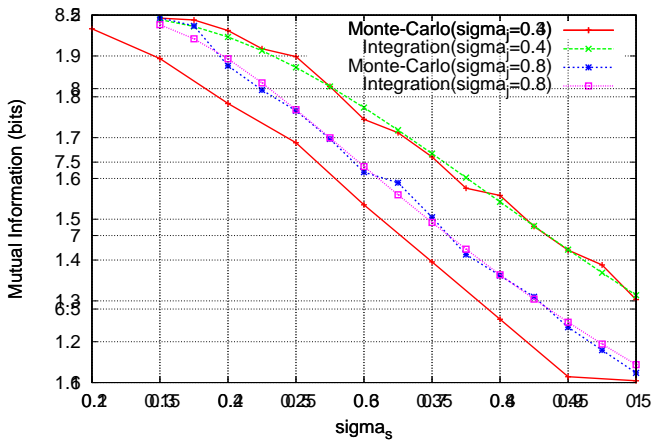}
\end{center}
\caption{Symmetric mutual information of two bit-cell model as a function of $\sigma_s$
(numerical integration and Monte Carlo method)}
\label{exp1}
\end{figure}

In this section, several numerical results obtained by the Monte Carlo method proposed in the previous section will be presented. 
Figure \ref{exp1} shows the plots of symmetric mutual information as a function of the standard deviation of the system noise $\sigma_s$. 
The curves of two cases such that $\sigma_j = 0.4$ and $\sigma_j = 0.8$ are shown in Fig.~\ref{exp1}. 
In this case,  numerical integration can be performed in order to obtain the symmetric mutual information 
because the number of bit-cells is small. We thus plot the results of numerical integration and the results of the Monte Carlo method in Fig.~\ref{exp1}.
The number of iterations is set to $t_{max} = 10000$ in the Monte Carlo simulations.
Although certain statistical fluctuations are observed for the curves obtained by the Monte Carlo simulations, the curves obtained by the two methods are in reasonable agreement. We can also observe a tendency whereby the symmetric mutual information decreases as $\sigma_j$ increases.

\begin{figure}[htbp]
\begin{center}
\includegraphics[scale=0.7]{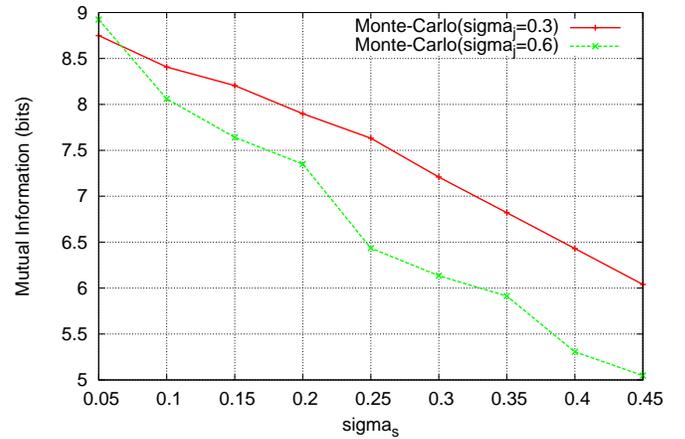}
\end{center}
\caption{Symmetric mutual information of simplified TDMR channel model as a function of $\sigma_s$ (9 bit-cells, $\alpha = 1.0, \beta = 0.5$, $t_{max} = 1000$)}
\label{exp2}
\end{figure}

Figure \ref{exp2} includes the curves of the symmetric mutual information of a simplified TDMR channel 
consisting of nine bit-cells. The interference coefficients $\alpha = 1.0, \beta = 0.5$ are 
assumed. The number of iterations is set to $t_{max} = 1000$.
In the case of $\sigma_j = 0.8$, the symmetric mutual information rate (mutual information divided by the number of bit-cells) approaches $2/3$ at $\sigma_s = 0.3$.

\section{Conclusion}

In order to capture the qualitative features of the writing and read-back processes of TDMR systems,
we proposed a simplified TDMR channel model. 
The simplicity of the proposed model enable us to derive the closed from of the conditional PDF representing 
the probabilistic nature of the channel. The conditional PDF is Gaussian distributed and is parameterized by 
the signal-dependent covariance matrix. Based on this conditional PDF, we developed a Monte Carlo method for approximating the symmetric mutual information. The symmetric mutual information is closely related to the areal density limit for a TDMR system without making any assumptions on coding or detection schemes. The numerical results presented herein suggest that a low-rate coding, such as 2/3 or 1/2, may be necessary when jitter-like noise becomes dominant.
The channel model is useful not only for estimation of the capacity of TDMR systems but also for the development 
of efficient detection algorithms and for the design of two-dimensional codes suitable for a TDMR channel. 
%Such applications remain to be developed.

\section*{Acknowledgement}

This work was supported by Storage Research Consortium.


\begin{thebibliography}{15}% 文献数が10未満の時 {9}

\bibitem{Khatami}  M. Khatami and B. Vacic, ``Constrained coding and detection for 
TDMR using generalized belief propagation,'' in proceedings of IEEE  International Conference on Communications (ICC),
pp. 3895--3901, 2014.

\bibitem{Chan}
K.S. Chan, R. Radhakrishnan, K. Eason, M.R. Elidrissi, J.J. Miles, B. Vasic, and A.R. Krishnan,
``Channel models and detectors for two-dimensional magnetic recording,''
pp.804--811, IEEE Transactions on Magnetics, vol. 46, no. 3, March, 2010.

\bibitem{Hwang}
E. Hwang, R.Negi, and B.V.K. Vijaya Kumar,
``Signal processing for near 10 Tbit/in$^2$ density in 
two-Dimensional magnetic recording (TDMR),''
pp.1813--1816, IEEE Transactions on Magnetics, vol. 46, no. 6, June, 2010.

\bibitem{Wood} R. Wood, M. Williams, A. Kavcic, and J. Miles,
``The feasibility of magnetic recording at 10 terabits per square inch on conventional media,''
pp.917--923, IEEE Transactions on Magnetics, vol. 45, no. 6, June, 2009.

\bibitem{Krishnan}
A. Krishnan, R. Radhakrishnan, B. Vasic, A. Kavcic, W. Ryan, and F. Erden,
``2-D magnetic recording: read channel modeling and detection,''
pp.3830--3826, IEEE Transactions on Magnetics, vol. 45, no. 10, Oct., 2009.

\bibitem{Cover}
T. M. Cover and J. A. Thomas,
``Elements of Information Theory,''  second ed.,  Wiley, 2006.

\end{thebibliography}
\end{document}